\documentclass[10pt, conference]{IEEEtran}
\IEEEoverridecommandlockouts 
\usepackage{graphicx}
\usepackage{color}
\usepackage{url}
\usepackage{algorithm}
\usepackage{algorithmic}
\usepackage{setspace}  
\usepackage{algorithm} 
\usepackage{algorithmic} 
\usepackage{multirow} 
\usepackage{amsmath} 
\usepackage{xcolor}
\usepackage{booktabs} 
\usepackage{threeparttable}
\usepackage{marvosym}
\usepackage{subfigure}
\usepackage{diagbox}
\usepackage[numbers,sort&compress]{natbib} 

\newcommand{\name}{LogStamp}

\ifCLASSINFOpdf

\else
 
\fi

\begin{document}

\title{LogStamp: Automatic Online Log Parsing Based on Sequence Labelling}


\author{
\IEEEauthorblockN{
Shimin Tao\IEEEauthorrefmark{4}, 
Weibin Meng\thanks{\Letter~Weibin Meng(mengweibin3@huawei.com) is corresponding author.}\IEEEauthorrefmark{4}\Letter, 
Yimeng Chen\IEEEauthorrefmark{4}, 
Yichen Zhu\IEEEauthorrefmark{3},
Ying Liu\IEEEauthorrefmark{2}
} 
\IEEEauthorblockN{
Chunning Du\IEEEauthorrefmark{5},
Tao Han\IEEEauthorrefmark{4},
Yongpeng Zhao\IEEEauthorrefmark{4},
Xiangguang Wang\IEEEauthorrefmark{4},
Hao Yang\IEEEauthorrefmark{4}
}

\IEEEauthorblockA{\IEEEauthorrefmark{4}Huawei,\IEEEauthorrefmark{3}University of Toronto}
\IEEEauthorblockA{\IEEEauthorrefmark{2}Tsinghua University}
\IEEEauthorblockA{\IEEEauthorrefmark{5}Beijing University of Posts and Telecommunications}

}

\maketitle

\begin{abstract}
Logs are one of the most critical data for service management.
It contains rich runtime information for both services and users. 
Since size of logs are often enormous in size and have free handwritten constructions, a typical log-based analysis needs to parse logs into structured format first. 
However, we observe that most existing log parsing methods cannot parse logs online, which is essential for online services.
In this paper, we present an automatic online log parsing method, name as \textit{LogStamp}.
We extensively evaluate LogStamp on five public datasets to demonstrate the effectiveness of our proposed method. The experiments show that our proposed method can achieve high accuracy with only a small portion of the training set. For example, it can achieve an average accuracy of 0.956 when using only 10\% of the data training.
\end{abstract}


\section{Introduction}

Logs are one of the most valuable data sources for large-scale services maintenance\cite{zhu2019tools,meng2020summarizing}, which report service runtime status and help operators to find trace workflows. 
Logs have been widely applied for a variety of service management and diagnostic tasks. Prior research has proposed automated approaches to analyze logs, such as status monitoring \cite{Khatuya2018ADELEAD}, anomaly detection \cite{loganomaly,zhu2021unilog},  failures prediction \cite{zhang2018prefix} and root cause analysis\cite{kobayashi2017mining}. 
The fast-emerging AIOps (Artificial Intelligence for IT Operations) solutions also utilize operation logs as their input data\cite{dai2020logram}.

\begin{figure}
      \begin{minipage}[h]{1.0\linewidth}
      \centering
      \includegraphics[width = 8.0 cm]{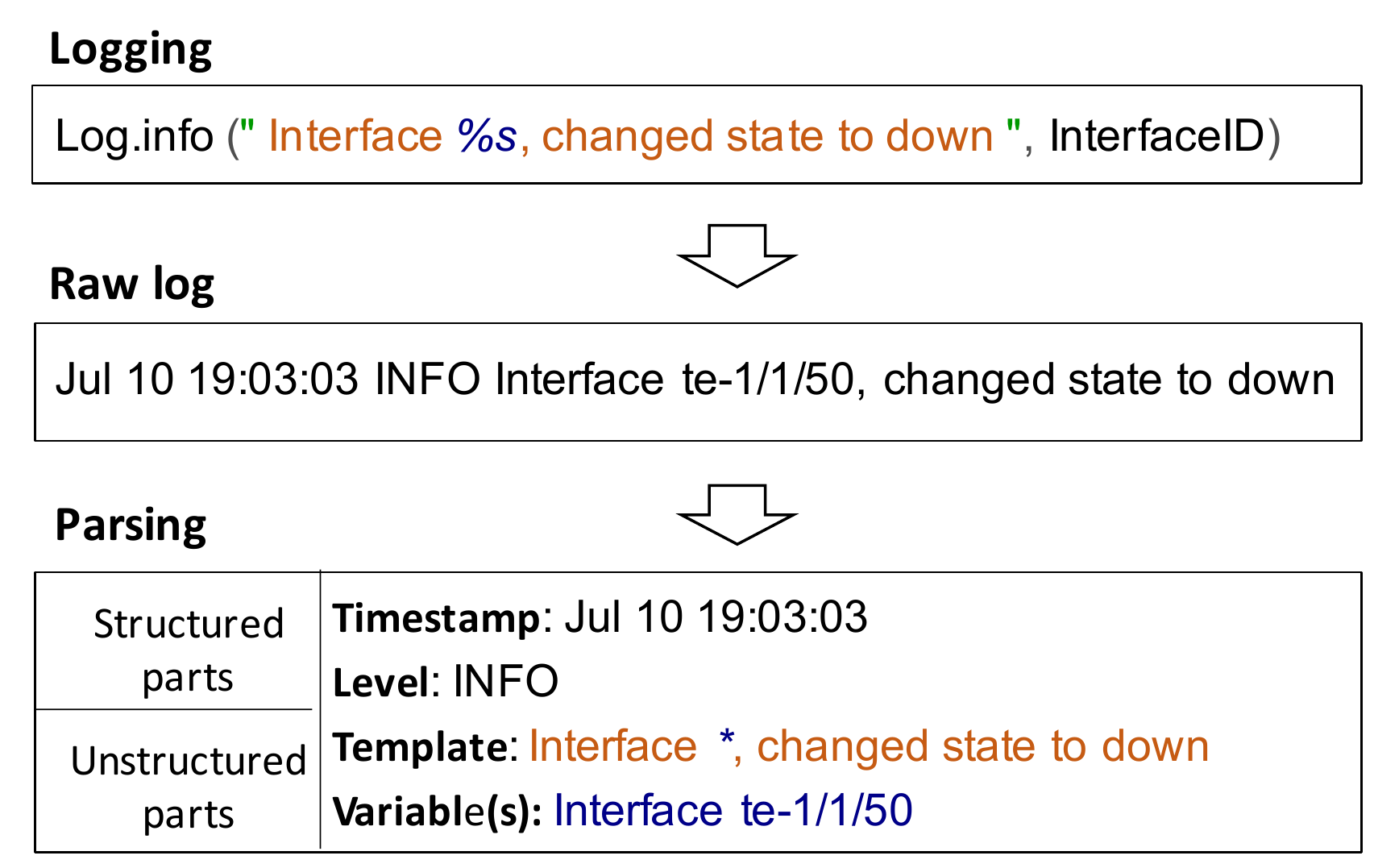}\\
      \end{minipage}
       \vspace{1 mm}
       \caption{An illustrative example of log parsing from source code of logging to structure log}\label{fig:logs}
    \vspace{1 mm}
\end{figure}  

Logs are designed by developers and generated by logging statements (\textit{e.g.}, \textit{printf}(), \textit{log.info}())) in the source code\cite{logparse}. 
As shown in Fig.\ref{fig:logs}, a logging function composes of log level (\textit{i.e.}, info), constant parts (\textit{i.e.}, ``Interface'' and ``change state to down''), and variables (\textit{i.e.}, “InterfaceID”). 
Service and system generate raw logs by printing an unstructured text that contains constant text and specified variables (\textit{e.g.}, ``te-1/1/50''). Usually, the constant parts sketch out the event and summarize it, and variables vary from one log to another of the same template.

Since logs are often extensive in size (\textit{e.g.}, Google and Facebook respectively generate 100 Petabyte and 10 Petabyte of log data per month\cite{logparse}) and have free handwritten constructions\cite{zhu2019tools}, log analysis remains a significant challenge. 
To address the challenge of the large size of logs, researchers propose approaches for log compression\cite{liu2019logzip}. However, most log compression approaches only aim to save storage space while not assisting when analyzing logs in practice. 
To address the challenge of log analysis, using rules (\textit{e.g.}, source code\cite{Xu2009Detecting} and regular expressions\cite{meng2021logclass}) is a simple yet effective approach. However, the source code is not always available, and designing regular expressions relies on domain knowledge, which cannot use in practice.
Therefore, automatic log parsing methods are getting attention. Researchers propose many approaches for automatically parsing raw logs into structured forms\cite{locke2021logassist}. The main aim of log parsing is to find templates (constant parts) from logs and replace variables with variable placeholders. 
Recently, many data-driven log parsing approaches have been proposed. There are multiple techniques, such as clustering \cite{lin2016log}, longest common subsequence\cite{spell}, frequent pattern mining\cite{ft-tree,spell},  heuristics parsing\cite{iplom} and others\cite{logparse}.  
However, log parsing still faces two challenges. 

Firstly, operators continuously conduct software/firmware upgrades on services/systems to introduce new features, fix bugs, or improve performance\cite{logparse}, which can generate new types of logs. 
Most of the existing approaches do not support online analysis. A small number of approaches (\textit{e.g.}, FT-tree\cite{ft-tree}, LogParse\cite{logparse}) that support online parsing also have some shortcomings, or they cannot handle new types of words, or they need to be combined with other log parsing algorithms to complete.
Therefore, \textit{Newly generated logs are difficult to process online }.

Besides, most existing log parsing approaches similar group logs and extracts templates for each group by keeping the same parts from logs and replacing different parts with placeholders. 
By default, log parsing is an unsupervised process. Parsers extract templates based on provided data instead of domain knowledge. Therefore, they only produce accurate results with sufficient historical log data. And, technically, the more data provided, the more accurate result they return.
However, when a brand new service goes online, there are usually not enough historical logs to generate accurate templates. Therefore, \textit{it's challenging to train a parsing model with small amounts of log data}.

To address the above challenges, we propose \name{}.
The key intuition is based on the following observations: When Operations reads the log, they mentally mark the words in the log to identify the template.
In \name{}, \textbf{we turn the log parsing problem into a sequence labelling problem} and find templates from logs online. 
\name{}'s contribution can be summarized as follows:

\begin{itemize}
    \item \name{} is an accurate online log parsing method. \name{} can parse logs one by one, and has extremely high accuracy. 
    \item \name{} can train an accurate log parsing model based on a small amount of log data, which ensures that it can analyze online logs. Experiments show that it can achieve an average accuracy of 0.956 when using only 10\% of the data training.
\end{itemize}

The rest of the paper is organized as follows: 
We discuss related works in Section \ref{sec:related} and propose our approach in Section \ref{sec:design}. 
The evaluation is shown in Section \ref{sec:evaluation}.
In Section \ref{sec:discussion}, we discuss \name{}'s limitations and future works.
Finally, we conclude our work in Section \ref{sec:conclusion}.
\label{sec:introduction}

\section{Related Work}\label{sec:related}

Logs play an important role in service management. 
Log parsing usually serves as the the first step towards automated log analysis~\cite{zhu2019tools}.


The most straightforward approach is to use rules to parse logs, such as regular expressions. The rule-based log parsing methods rely on handcrafted rules provided by domain knowledge.
Though straightforward, this kind of method requires a deep understanding of the logs, and a lot of manual efforts are needed to write different rules for different kinds of logs, which is not general.
Commercial log analytic platforms (\textit{e.g.}, Splunk, ELK, Logentries) also allow operators to efficiently manage and analyze large-scale logs by pre-define rules. But they are only applicable to certain types of logs and are not universal.

Utilizing source code can parsing logs accurate. For example, \cite{Xu2009Detecting} employs source code to extract log templates for system problem detection. However, the source code is not always available, especially for commercial services.

To achieve the goal of automated log parsing, many data-driven approaches have been proposed. There are many categories of log parsing~\cite{zhu2019tools,zhu2021student}. 
The first category is cluster-based approaches, which log template forms a natural pattern of a group of log messages. From this view, log parsing can be modeled as a clustering problem, such as LogSig \cite{logsig}.
Next is longest common subsequence. For example, Spell \cite{spell} uses the longest common subsequence algorithm to parse logs in a stream. 
Iterative partitioning is used in IPLoM~\cite{iplom}. 
Some methods use heuristics to extract templates. As opposed to general text data, log messages have some unique characteristics. 
Consequently, Drain \cite{drain} propose heuristics-based log parsing methods. 
The next category is frequent item mining, which is straightforward. Tokens, which regularly appear together in different log entries, are built into frequent itemsets. 
The parser obtains templates by looking up those itemsets.
Log templates can be seen as a set of constant tokens that frequently occur in logs, such as FT-tree~\cite{ft-tree}. 
The final category is combined approaches. LogParse~\cite{logparse} combines existing unsupervised log parsing approaches and supervised machine learning approaches to generate templates for online logs. 
The idea of LogParse is similar to our paper. However, it's a pipeline workflow, which will be affected by the accuracy of traditional log parsing approaches because most log parsing approaches cannot achieve high accuracy based on a small number of logs.

\section{Design}\label{sec:design}
\begin{figure*}
      \begin{minipage}[h]{1.0\linewidth}
      \centering
      \includegraphics[width = 15 cm]{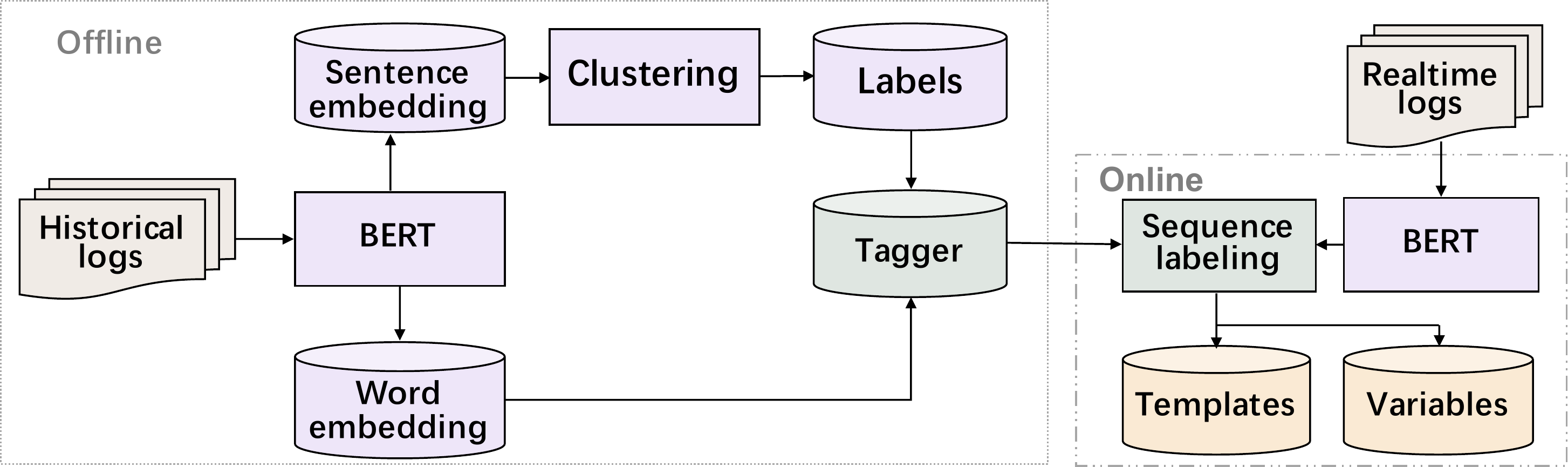}\\
      \end{minipage}
       \vspace{1 mm}
       \caption{The workflow of \name{}}\label{fig:design}
    \vspace{1 mm}
\end{figure*}

In this section, we introduce the overall framework of our proposed LogStamp. The overview of the framework is shown in Fig.~\ref{fig:design}. We first present the offline part in our workflow in Section~\ref{sec:online}, then we will describe the online part in detail in Section~\ref{sec:offline}. 

\subsection{Offline workflow}
\label{sec:offline}
\noindent
Given a set of historical logs, our goal is to build a tagger to identify if the incoming log is a template or variable. Previous works~\cite{logparse} use the template extraction method to obtain the templates from the logs. Then, a word classifier (i.e., SVM classifier) is adopted to label each word in the logs. There are two drawbacks to such methods. First, the accuracy of labeling depends on the quality of the extracted template. If the template extraction method fails to separate the templates from the raw logs, the pseudo label assign by the classifier would be meaningless and cause failure on log parsing. Secondly, prior works only utilize templates to train the word classifier. Generally, log data contains critical information on both word-level and sentence-level (i.e., sentence order). For example, in log anomaly detection task, a common way to detect the anomalous logs in the log data is to see if the log orders are correct. If we have received a log says that "\textit{Vlan-Interfenerce ae, change state to up}" and no message of "\textit{Vlan-Interfenerce ae, change state to down}" is followed in a certain time period, we will recognize such log as anomalous log. And because the word-level embedding only focus on the single word, it fails to effective parse such logs .Therefore, learning logs feature from the sentence level is important.

In this paper, we introduce a coarse to the fine framework to generate accurate pseudo labels. First of all, a pretrained bidirectional transformer is adopted to extract the feature representation of log data. Because the structure of raw logs is different from the natural language, we need to finetune the BERT~\cite{devlin2018bert} using our data. Note that finetuning the BERT does not need any label. Then we use a dual-path framework to get both coarse level embedding and fine level embedding. On the coarse level, we expect the sentence embedding can reflect the nature of different logs. For instance, the above example of two logs have similar structure, and most of the words in these two logs are the same. However, the meaning of these two logs are completely different. The coarse level feature learns the inherent relations between the words, thus output two embeddings with distant similarity. The sentence embedding can be further group into number of clusters.

In general, one can exploit any clustering algorithm that can split the sentence into clusters according to their embedding features. Our approach is to use DBSCAN~\cite{ester1996density}. After we get the clusters, we count the frequency of word appearance in each clustering. We mark it as template if the number of appearance is larger than the threshold, variables otherwise. As such, we obtain the labels for each word.

For the fine-grained level, we used the finetuned BERT to output word embeddings. For each word embedding, we have its corresponding label from the step above. Given a set of word embeddings and word labels, we can train a classifier that serves as a tagger. As we trained via a deep neural network, this tagger can accurately parse the logs without the interference introduced by the wrong pseudo labels. 

\subsection{Online workflow}\label{sec:online}
In real-time systems, systems may generate new log templates online; therefore, building a robust online workflow is critical for real scenario deployment. Our online workflow is simple. Given real-time logs, which can be either a piece of logging information or a set of new logs, we reuse the BERT model to extract the word embedding from the new logs. Then the tagger that is trained in the offline stage will predict a label for the logs. As a result, we can immediately know whether the specific words are templates or variables. We will show that our online framework is simple yet surprisingly effective under most of the circumstance.

\section{Evaluation}\label{sec:evaluation}

In this section, we evaluate our approach using public log datasets and aim to answer the following research questions:
\begin{itemize}
    \item RQ1: How effective is \name{} in log parsing?
    \item RQ2: Can \name{} achieve accurate results based on a small amount of log data? 
    \item RQ3: How much can the BERT and tagger contribute to the overall performance?
\end{itemize}



\subsection{Experiment Setting}\label{sec:setting}

In this section, we evaluate the performance of \name{}.
The datasets, baselines, evaluation metrics and experimental setup of the experiments are as follows.

\begin{figure*}
      \begin{minipage}[h]{1.0\linewidth}
      \centering
      \includegraphics[width = 16 cm]{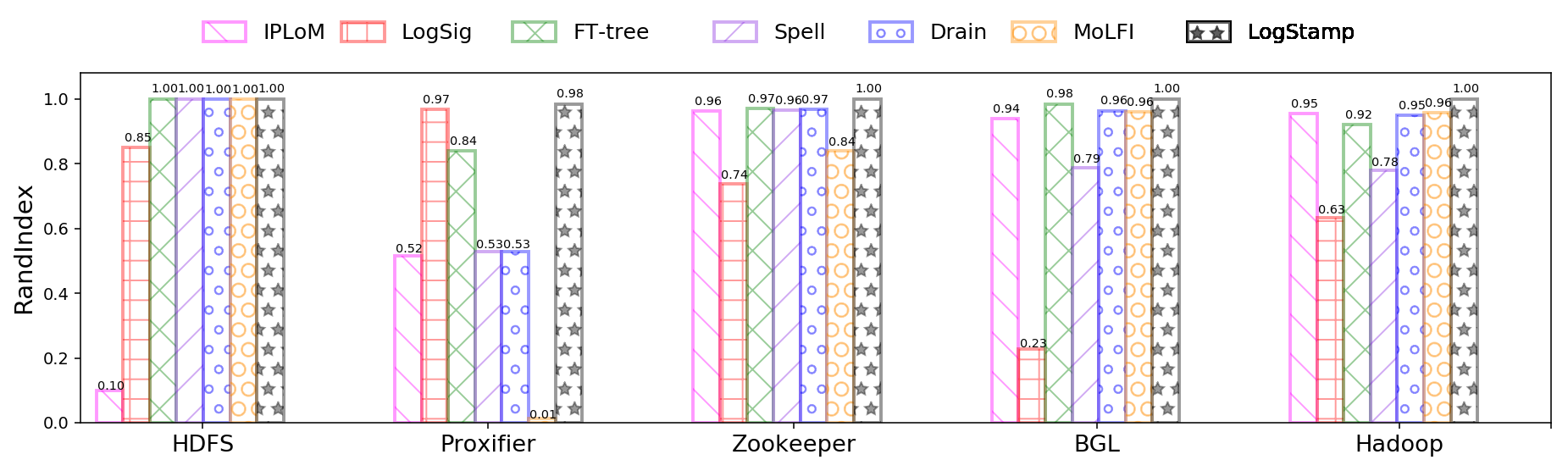}\\
      \end{minipage}
       \vspace{1 mm}
       \caption{Comparison of the accuracy of offline log parsing between LogStamp and six baselines when they are trained by all offline logs}\label{fig:offline_match}
    \vspace{1 mm}
\end{figure*}

\begin{figure*}
      \begin{minipage}[h]{1.0\linewidth}
      \centering
      \includegraphics[width = 16 cm]{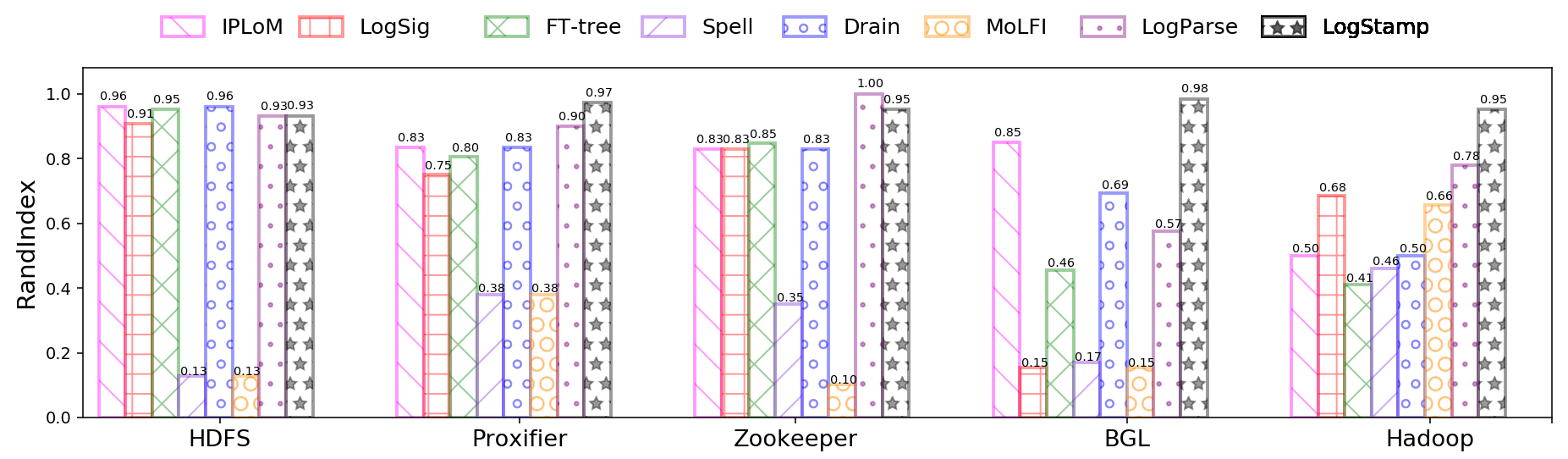}\\
      \end{minipage}
       \vspace{1 mm}
       \caption{Comparison of the accuracy of online log parsing between LogStamp and seven baselines when they are trained by 10\% offline logs}\label{fig:online_10train}
    \vspace{1 mm}
\end{figure*}

\subsubsection{Datasets}\label{sec:datasets}
We conduct experiments over five public log datasets from distributed systems, which are BGL \cite{zhu2019tools}, HDFS \cite{xu2009largescale}, ZooKeeper \cite{he2016experience}, Proxifier \cite{zhu2019tools} and Hadoop \cite{lin2016log}.
The detailed information of these datasets is listed in Table~\ref{tab:datasets}. 
For each dataset, \cite{zhu2019tools} sampled logs and manually labelled each log's template, which serves as the ground truth for our evaluation.


\begin{table} 
\caption{Detail of the datasets}\label{tab:datasets}
\centering
\renewcommand\tabcolsep{2.75 pt} 
\begin{tabular}{cc}
\toprule
Datasets&Description\\ 
\midrule
HDFS&Hadoop~distributed~file~system\\ 
Proxifier&Proxifier software \\
ZooKeeper&ZooKeeper service \\
BGL&Blue~Gene/L~supercomputer\\ 
Hadoop&Hadoop MapReduce job \\
\toprule
\end{tabular}
\end{table}

\subsubsection{Baselines}\label{sec:baselines}

To demonstrate the performance of \name{}, we have implemented five template extraction methods: FT-tree \cite{ft-tree}, Drain \cite{drain}, Spell \cite{spell}, LogSig \cite{logsig}, LogParse~\cite{logparse}, MoLFI~\cite{molfi} and IPLoM \cite{iplom} . The parameters of these methods are all set best for accuracy. 
LogParse~\cite{logparse} can incorporate any existing template extraction method, our paper utilized Spell to initialize LogParse. 

For BERT, we use three versions of BERT, \textit{i.e.}, BERT-base, BERT-tiny and BERT-small.
For tagger in \name{}, we compare the performances of GCN, CNN, LSTM and RNN.

\subsubsection{Evaluation Metrics}\label{sec:metrics}
We apply $RandIndex$~\cite{randindex} to quantitatively evaluate the accuracy of template extraction.
$RandIndex$ is a popular method for evaluating the similarity between two data clustering techniques or multi-class classifications.  
What's more, $RandIndex$ is applied to evaluating existing template extraction methods in the literature, such as in \cite{logparse}.

For each template extraction method, we evaluate its accuracy by calculating the $RandIndex$ between the manual classification results and the templates learned by it.
Specifically, among the template learning results of a specific method, we randomly select two logs, \textit{i.e.}, $x$ and $y$, and define $TP, TN, FP, FN$ as follows.
$TP$: $x$ and $y$ are manually classified into the same cluster and they have the same template; 
$TN$: $x$ and $y$ are manually classified into different clusters and they have different templates; 
$FP$: $x$ and $y$ are manually classified into different clusters and they have the same template; 
$FN$: $x$ and $y$ are manually classified into the same cluster and they have different templates. 
Then $RandIndex$ can be calculated using the above terms as follows: $RandIndex = \frac{TP+TN}{TP+TN+FP+FN}$.

\subsubsection{Experimental Setup}\label{sec:setup}
We conduct experiments on a Linux server with Intel Xeon 2.40 GHz CPU and 64G memory. 

\subsection{Evaluation Results}\label{sec:results}

\textbf{1) RQ1: How effective is \name{} in log parsing?}

We first compare accuracies of existing log parsing methods\footnote{LogParse's offline step utilized other log parsing methods. It doesn't have its own offline parsing.} and LogStamp when extract template from historical logs. The comparison results are shown in Fig.~\ref{fig:offline_match}.
We find that most log parsing methods are highly accurate in extract templates from historical logs.
However, the accuracy of existing parsers is not always consistent. In other words, the selection of log data impacts the parsing accuracy~\cite{zhu2019tools}. Parsers may have a good evaluation result with up to 90\% of accuracy and an unacceptable bad outcome down to 50\% depending on different input datasets (\textit{e.g.}, Proxifier).
Meanwhile, we find LogStamp still achieves high accuracy (the average accuracy is more than 0.999) on different datasets. Therefore, we can directly use the label results to train a tagger.

To demonstrate the performance of \name{} in supporting online parsing and simulate the launch of new services, for each dataset, we apply each log parsing method to extract templates from 10\% of their logs.
Fig. \ref{fig:online_10train} shows the comparative results.
LogStamp achieves the best performance.
Specifically, the accuracy of LogStamp on each dataset is 0.956.

\textbf{2) RQ2: Can \name{} achieve accurate results based on a small amount of log data?}

\begin{figure}
      \begin{minipage}[h]{1.0\linewidth}
      \centering
      \includegraphics[width = 9.4 cm]{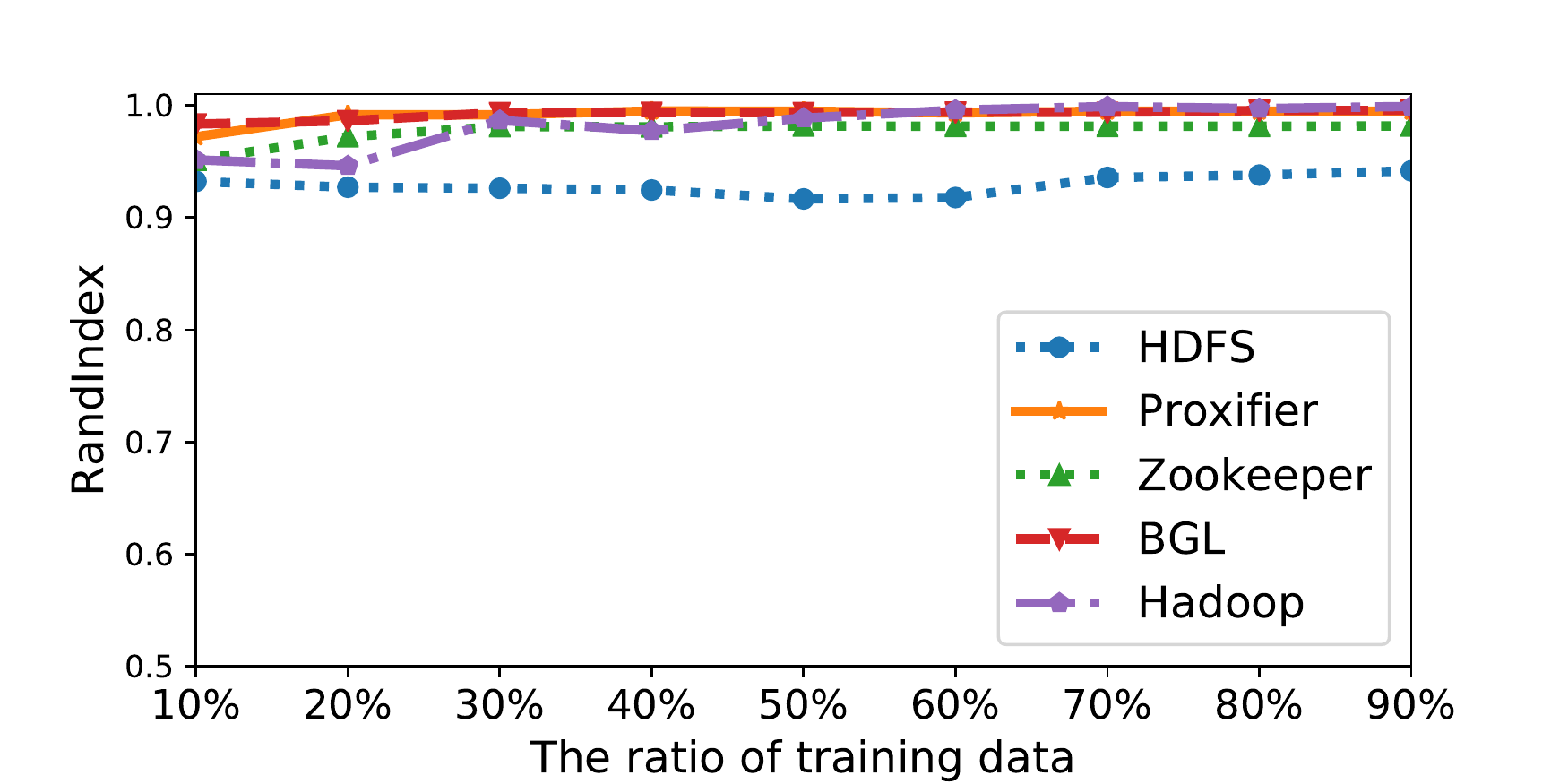}\\
      \end{minipage}
       \vspace{1 mm}
       \caption{The log parsing accuracy of \name{} as the ratio of training data changes}\label{fig:stable}
    \vspace{1 mm}
\end{figure}

As shown in \cite{zhu2019tools}, the accuracy of existing parsers is not always consistent, both for the datasets and the percentage of training data.
To demonstrate how stable \name{} is to the scale of training data, Fig. \ref{fig:stable} shows the log parsing accuracy of \name{} on the five datasets, as the percentage of training data increases from 10\% to 90\%, respectively.
The results show that \name{} is stable to different scales of training logs and can achieve high log parsing accuracy when trained based on a small scale of training data.

\textbf{3) RQ3: How much can the BERT and tagger contribute to the overall performance?}

LogStamp incorporates two modules: BERT and taggers. In this RQ, we evaluate the effectiveness of different version of each module. 
Firstly, we compare LogStamp with BERT-base, BERT-small and BERT-tiny. Table~\ref{tab:bert-offline} and Table~\ref{tab:bert-online} show the performance of LogStamp in the offline stage and the online stage, respectively. We find that three versions of BERT achieve similar performance, which means that LogStamp doesn't need to spend time adjusting the effect of BERT.
Then, in Table~\ref{tab:taggers}, we compare LogStamp with different taggers, \textit{i.e.}, GCN, RNN, LSTM and CNN.
We find that LSTM achieves the best performance on all datasets. 
Because LSTM is more suitable to natural language processing, and sequence labelling is a problem in natural language processing.


\begin{table} 
\caption{Offline accuracy of LogStamp with different BERT versions}\label{tab:bert-offline}
\centering
\renewcommand\tabcolsep{2.75 pt} 
\begin{tabular}{cccccc}
\toprule
\multirow{2}*{Methods}&\multicolumn{5}{c}{Datasets}\\ 
~&HDFS&Proxifier&Zookeeper&BGL&Hadoop\\
\midrule
BERT-tiny&0.9999&0.9356&0.9998&0.9950&0.9988\\ 
BERT-base&0.9999&0.9836&0.9998&0.9994&0.9987 \\
BERT-small&0.9999&0.9840&0.9998&0.9979&0.9988 \\
\toprule
\end{tabular}
\end{table}

\begin{table} 
\caption{Online accuracy of LogStamp with different BERT versions}\label{tab:bert-online}
\centering
\renewcommand\tabcolsep{2.75 pt} 
\begin{tabular}{cccccc}
\toprule
\multirow{2}*{Methods}&\multicolumn{5}{c}{Datasets}\\ 
~&HDFS&Proxifier&Zookeeper&BGL&Hadoop\\
\midrule
BERT-tiny&0.8888&0.9042&0.9906&0.9788&0.9762\\ 
BERT-base&0.8798&0.9141&0.9760&0.9816&0.9637 \\
BERT-small&0.9147&0.8820&0.9851&0.9586&0.9752 \\
\toprule
\end{tabular}
\end{table}

\begin{table} 
\caption{Online Accuracy of LogStamp with different taggers}\label{tab:taggers}
\centering
\renewcommand\tabcolsep{2.75 pt} 
\begin{tabular}{cccccc}
\toprule
\multirow{2}*{Methods}&\multicolumn{5}{c}{Datasets}\\ 
~&HDFS&Proxifier&Zookeeper&BGL&Hadoop\\
\midrule
GCN&0.8888&0.9042&0.9906&0.9788&0.9762\\ 
RNN&0.9822&0.9180&0.9790&0.9978&0.9962 \\
LSTM&0.9949&0.9998&0.9998&0.9996&0.9974 \\
CNN&0.9921&0.9164&0.9998&0.9996&0.9974 \\
\toprule
\end{tabular}
\end{table}


\section{Discussion and Future Work}\label{sec:discussion}
Thanks to BERT for its powerful ability to capture both sentence embedding of log sentences for clustering and word embedding for distinguishing between templates and variables in log. However, during experiments, it is observed that syntactic structure and semantic information contained in log sentences often vary considerably compared to those sentences used to train BERT. One deduction is that if the BERT model is fine-tuned on log datasets with masked language modeling, it might better understand log sentences and thus have higher accuracy in offline and online log parsing. Yet, the experiment result does not prove the deduction to be correct. By fine-tuning the BERT model with log sentences of each system for 1-3 epochs, the online clustering rand index does not seem to be steadily improved.

We will continue to study how to apply better pre-trained language models to log template extractions in our future work. More abundant logs will be used to fine-tune BERT or to train a BERT from the beginning instead of directly loading weights of model pre-trained with dissimilar vocabularies, \textit{e.g.}, from Wikipedia or books. Besides, as log sentences usually have a more unified structure, we will also attempt to design a more concise model structure based on BERT to achieve higher efficiency in online log parsing to deal with higher concurrency.

\section{Conclusion}\label{sec:conclusion}
In this paper, we propose LogStamp, an online log parsing approach. 
Different from the prior log parsing approach, LogStamp takes semantic into consideration and turns the log parsing problem into a sequence labelling problem. LogStamp supports a training model based on a small number of historical logs. Experimental results on public log datasets have validated the accuracy and stability of LogStamp.


\newpage
\bibliographystyle{unsrt} 
\bibliography{references}

\end{document}